\documentclass[12pt]{article}
\let\section=\subsection     \let\subsection=\subsubsection                
\usepackage{graphicx}

\begin{document}
\begin{center}
   {\large \bf Hadron structure in manifestly Lorentz-invariant}\\[2mm]
   {\large \bf baryon chiral perturbation theory}\\[5mm]
   T.~Fuchs, M.~R.~Schindler, J.~Gegelia and S.~Scherer\\[5mm]
   {\small \it  Institut f\"ur Kernphysik, Johannes Gutenberg-Universit\"at 
   Mainz \\
   J.~J.~Becher Weg 45, D-55099 Mainz, Germany \\[8mm] }
\end{center}

\begin{abstract}\noindent
   We briefly outline the so-called extended on-mass-shell renormalization 
scheme for manifestly Lorentz-invariant baryon chiral perturbation theory 
which provides a simple and consistent power counting for renormalized 
diagrams.
   We comment on the role of chiral symmetry in the renormalization program
and discuss as applications the mass and the electromagnetic 
form factors of the nucleon.
\end{abstract}

\section{Introduction}
   Mesonic chiral perturbation theory (ChPT)
\cite{Weinberg:1978kz,Gasser:1983yg}
has been tremendously successful and may be considered as
a full-grown and mature area of low-energy particle physics 
(for a recent review, see, e.g., Ref.\ \cite{Scherer:2002tk}).
    The prerequisite for an effective field theory program 
is (a) a knowledge of the most general 
effective Lagrangian and (b) an expansion scheme for observables in terms 
of a consistent power counting method \cite{Weinberg:1978kz}.
   In the mesonic sector the Lagrangian is known up to and including
${\cal O}(q^6)$ in the momentum and quark mass expansion 
\cite{p6}.
   The combination of dimensional regularization with the modified
minimal subtraction scheme of ChPT \cite{Gasser:1983yg} leads to
a straightforward correspondence between the loop expansion and the 
chiral expansion in terms of momenta and quark masses at a fixed ratio,
and thus provides a consistent power counting for renormalized quantities.
   In the extension to the one-nucleon sector \cite{Gasser:1988rb} the 
correspondence between the loop expansion and the chiral expansion,
at first sight, seems to be lost: higher-loop 
diagrams can contribute to terms as low as ${\cal O}(q^2)$
\cite{Gasser:1988rb}.
   This problem has been eluded in the framework of the 
heavy-baryon formulation of ChPT \cite{HB}, 
resulting in a power counting analogous to the mesonic sector.
   The price one pays consists of giving up manifest Lorentz invariance
of the Lagrangian. 
   In addition, at higher orders in the chiral expansion, the 
expressions due to $1/m$ corrections of the Lagrangian become increasingly 
complicated \cite{HBLag}.  
   Finally, not all of the scattering amplitudes, evaluated perturbatively
in the heavy-baryon framework, show the correct analytical behavior in the 
low-energy region.
   In the following we will outline some recent developments in devising
a renormalization scheme leading to a simple and consistent power counting 
for the renormalized diagrams of a manifestly Lorentz-invariant approach.

\section{Manifestly Lorentz-Invariant Baryon Chiral Perturbation Theory
and EOMS Scheme}
   In order to illustrate the issue of power counting, let us consider
the lowest-order $\pi N$ Lagrangian \cite{Gasser:1988rb},
expressed in terms of bare fields and parameters denoted by subscripts 0,
\begin{equation} 
\label{lpin1}
{\cal L}_{\pi N}^{(1)}=\bar \Psi_0 \left( i\gamma_\mu 
\partial^\mu -m_0 -\frac{1}{2}\frac{{\stackrel{\circ}{g_{A}}}_0}{F_0} 
\gamma_\mu 
\gamma_5 \tau^a \partial^\mu \pi^a_0\right) \Psi_0 +\cdots,
\end{equation} 
   where $\Psi_0$ and $\vec{\pi}_0$ denote a doublet and a triplet of bare 
nucleon and pion fields, respectively.
   After renormalization, $m$, $\stackrel{\circ}{g_A}$, and $F$ refer to the 
chiral limit of the physical nucleon mass, the axial-vector coupling constant,
and the pion-decay constant, respectively. 
   The most general effective Lagrangian of the interaction of Goldstone 
bosons with nucleons consists of a string of terms 
\begin{displaymath}
{\cal L}_{\pi N}={\cal L}_{\pi N}^{(1)}+{\cal L}_{\pi N}^{(2)}+\cdots,
\end{displaymath}
where the superscripts refer to the order in the derivative and
quark-mass expansion \cite{Gasser:1988rb}.
   In addition, one needs the most general effective Lagrangian of the
mesonic sector \cite{Gasser:1983yg,p6}
\begin{displaymath}
{\cal L}_{\pi}={\cal L}_2+ {\cal L}_4 +\cdots,
\end{displaymath}
containing only even powers in the chiral expansion. 

   The aim is to devise a renormalization procedure generating, after
renormalization, the following power counting:
   a loop integration in $n$ dimensions counts as $q^n$, 
pion and fermion propagators count as $q^{-2}$ and 
$q^{-1}$, respectively, vertices derived from ${\cal L}_{2k}$ and 
${\cal L}_{\pi N}^{(k)}$ count as $q^{2k}$ and $q^k$, respectively.
   Here, $q$ generically denotes a small expansion parameter such as,
e.g., the pion mass.
   In total this yields for the power $D$ of a diagram in the 
one-nucleon sector the standard formula 
\begin{eqnarray}
\label{dimension1}
D&=&n N_L - 2 I_\pi - I_N +\sum_{k=1}^\infty 2k N^\pi_{2k}
+\sum_{k=1}^\infty k N_k^N,
\end{eqnarray}
   where $N_L$, $I_\pi$, $I_N$, $N_{2k}^\pi$, and $N_k^N$ denote the
number of independent loop momenta, internal pion lines, internal nucleon 
lines, vertices originating from ${\cal L}_{2k}$,  
and vertices originating from ${\cal L}_{\pi N}^{(k)}$, respectively.   

\begin{figure}
\begin{center}
\includegraphics[height=3cm]{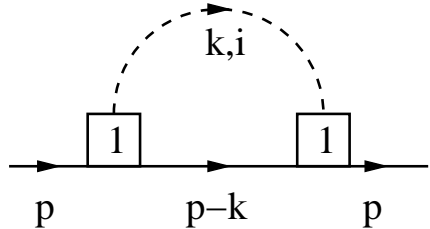}
\caption{One-loop contribution to the nucleon self-energy.
The number 1 in the interaction blobs refers to ${\cal L}_{\pi N}^{(1)}$.}
\label{fig:nucleonselfenergypionloop}  
\end{center}
\end{figure}

   As an example, let us consider the one-loop contribution of 
Fig.\ \ref{fig:nucleonselfenergypionloop} to the nucleon self-energy.
   According to Eq.\ (\ref{dimension1}), after renormalization, we would like
to have the order
\begin{equation}
\label{dexample}
D=n\cdot 1-2\cdot 1-1+1\cdot 2=n-1.
\end{equation}
   Applying the $\widetilde{\rm MS}$ renormalization scheme of ChPT
\cite{Gasser:1983yg,Gasser:1988rb}---indicated 
by ``r''---one obtains 
\begin{displaymath} 
\Sigma_{\rm loop}^r=-\frac{3 g_{Ar}^2}{4 F_r^2}\left[ 
-\frac{M^2}{16\pi^2}(p\hspace{-.4em}/\hspace{.1em}+m)
+\cdots\right]=
{\cal O}(q^2),
\end{displaymath}
where $M^2$ is the lowest-order expression for the squared pion mass.
   In other words, the $\widetilde{\rm MS}$-renormalized result does not 
produce the desired low-energy behavior of Eq.\ (\ref{dexample}).
   This finding has widely been interpreted as the absence of a systematic
power counting in the relativistic formulation of ChPT.

   Recently, several methods have been suggested to obtain a consistent
power counting in a manifestly Lorentz-invariant approach 
\cite{Tang:1996ca,Becher:1999he,Gegelia:1999gf,Lutz:1999yr,%
Fuchs:2003qc,Gegelia:1999qt,Fuchs:2003sh,Fuchs:2003kq}. 
   Here, we will concentrate on the so-called extended on-mass-shell (EOMS) 
renormalization scheme \cite{Fuchs:2003qc}.
   The central idea of the EOMS scheme consists of performing additional 
subtractions beyond the $\widetilde{\rm MS}$ scheme.
   Since the terms violating the power counting are analytic in small
quantities, they can be absorbed by counterterm contributions.
   We will illustrate our approach in terms of the integral 
\begin{displaymath}
H(p^2,m^2;n)= \int \frac{d^n k}{(2\pi)^n}
\frac{i}{[(k-p)^2-m^2+i0^+][k^2+i0^+]},
\end{displaymath}
where $\Delta=(p^2-m^2)/m^2={\cal O}(q)$ is a small quantity.
   We want the (renormalized) integral to be of the order $D=n-1-2=n-3$.
   Applying the dimensional counting analysis of Ref.\ \cite{Gegelia:zz}
(for an illustration, see the appendix of Ref.\ \cite{Schindler:2003je}),
the result of the integration is of the form \cite{Fuchs:2003qc}
\begin{displaymath}
H\sim F(n,\Delta)+\Delta^{n-3}G(n,\Delta),
\end{displaymath}
where $F$ and $G$ are hypergeometric functions and are analytic in $\Delta$ for
any $n$.
   Hence, the part containing $G$ for noninteger $n$ is proportional to
a noninteger power of $\Delta$ and satisfies the power counting.
   On the other hand $F$ violates the power counting.
   The crucial observation is that the part proportional to $F$ can be 
obtained by {\em first} expanding the integrand in small quantities and 
{\em then} performing the integration for each term \cite{Gegelia:zz}.
   This observation suggests the following procedure: expand the integrand in 
small quantities and subtract those (integrated) terms whose order is 
smaller than suggested by the power counting.
   In the present case, the subtraction term reads
\begin{displaymath}
H^{\rm subtr}=\int \frac{d^n k}{(2\pi)^n}\left.
\frac{i}{[k^2-2p\cdot k +i0^+][k^2+i0^+]}\right|_{p^2=m^2}
\end{displaymath}
and the renormalized integral is written as $
H^R=H-H^{\rm subtr}={\cal O}(q^3)
$
as $n\to 4$.
   In the infrared renormalization scheme of Becher and Leutwyler
\cite{Becher:1999he}, one would keep the contribution
proportional to $G$ (with subtracted divergences when $n$ approaches
4) and completely drop the $F$ term.

\section{The EOMS Scheme and Chiral Symmetry}   
   In the chiral limit of massless $u$ and $d$ quarks,
the QCD Lagrangian has a {\em global}
$\mbox{SU}(2)_L\times\mbox{SU(2)}_R\times\mbox{U}(1)_V$ symmetry.
   As a consequence of this symmetry, Green functions involving the Noether
currents are constrained by Ward identities.
   The Green functions may most efficiently be combined in a generating
functional through a coupling of the quark bilinears to external
fields \cite{Gasser:1983yg}.
   In the framework of chiral perturbation theory, the generating
functional is calculated
by means of the most general
effective mesonic and $\pi N$ Lagrangians.
   By construction, the effective Lagrangian of the relativistic formulation
is manifestly chirally invariant under {\em local}
$\mbox{SU}(2)_L\times\mbox{SU(2)}_R\times\mbox{U}(1)_V$
transformations provided the external fields are transformed
accordingly \cite{Gasser:1983yg}.
   The {\em local} invariance of the Lagrangian guarantees that the chiral
Ward identities of QCD
(as well as their symmetry-breaking pattern)
are encoded in the generating functional which is now
determined through the {\em effective} field theory.

   In the following we will briefly discuss the consequences of chiral 
symmetry for the renormalization program (for a similar discussion, see
Ref.\ \cite{Becher:1999he}).
   The tree graphs calculated in terms of the effective Lagrangian
separately satisfy the Ward identities.
   Dimensional regularization is known to respect the symmetry relations
induced by chiral symmetry for arbitrary $n$ so that the
corresponding regularized loop diagrams also satisfy the Ward
identities.
   The one-loop diagrams may be divided into two parts: the first
part is proportional to noninteger power(s) of the small
expansion parameter(s) and the second part is analytic.
   The nonanalytic parts cannot be altered by changing the renormalization
prescription and thus necessarily satisfy the Ward identities
independently from the analytic parts. The analytic parts satisfy
the Ward identities order by order in small expansion parameters.
    In our renormalization procedure we subtract all terms of the
expansion of the analytic parts of the one-loop diagrams that
violate the power counting. 
   These subtraction terms satisfy the Ward identities order by order.
   Hence the renormalized diagrams also respect the Ward identities.

   For multi-loop diagrams the procedure is analogous
albeit technically more complicated.
   For example, two-loop diagrams may contain parts which are
nonanalytic in the expansion parameter(s) and which cannot be
altered by the renormalization condition.
   These parts do not violate the power counting and
satisfy the Ward identities separately.
   However, there may also be contributions which are nonanalytic but depend
on the renormalization condition for the one-loop sub-diagrams.
   If the finite parts of the counterterms are fixed so that the
power counting is satisfied at the one-loop level, then these
parts of two-loop diagrams, combined with contributions from
counterterm diagrams renormalizing one-loop sub-diagrams, satisfy
power counting (see Ref.\ \cite{Schindler:2003je}).
   As long as the renormalization of the one-loop diagrams respects chiral
symmetry, the above second type of nonanalytic parts also
satisfies the Ward identities.
   Finally, the third part is analytic and can be altered by counterterms.
   Starting from here, the argument is as for the one-loop case.
   This (standard) procedure of renormalization is then performed iteratively
for diagrams with an increasing number of loops.

\section{Applications}
   Let us first discuss the result for the mass of the nucleon
at ${\cal O}(q^4)$ \cite{Fuchs:2003qc},
\begin{equation}
\label{mnoq4} 
m_N=m+k_1 M^2+k_2 M^3+k_3 M^4\ln\left(\frac{M}{m}\right)+k_4 M^4
+{\cal O}(M^5), 
\end{equation}
\label{applications}
\begin{center}
\includegraphics[width=8cm]{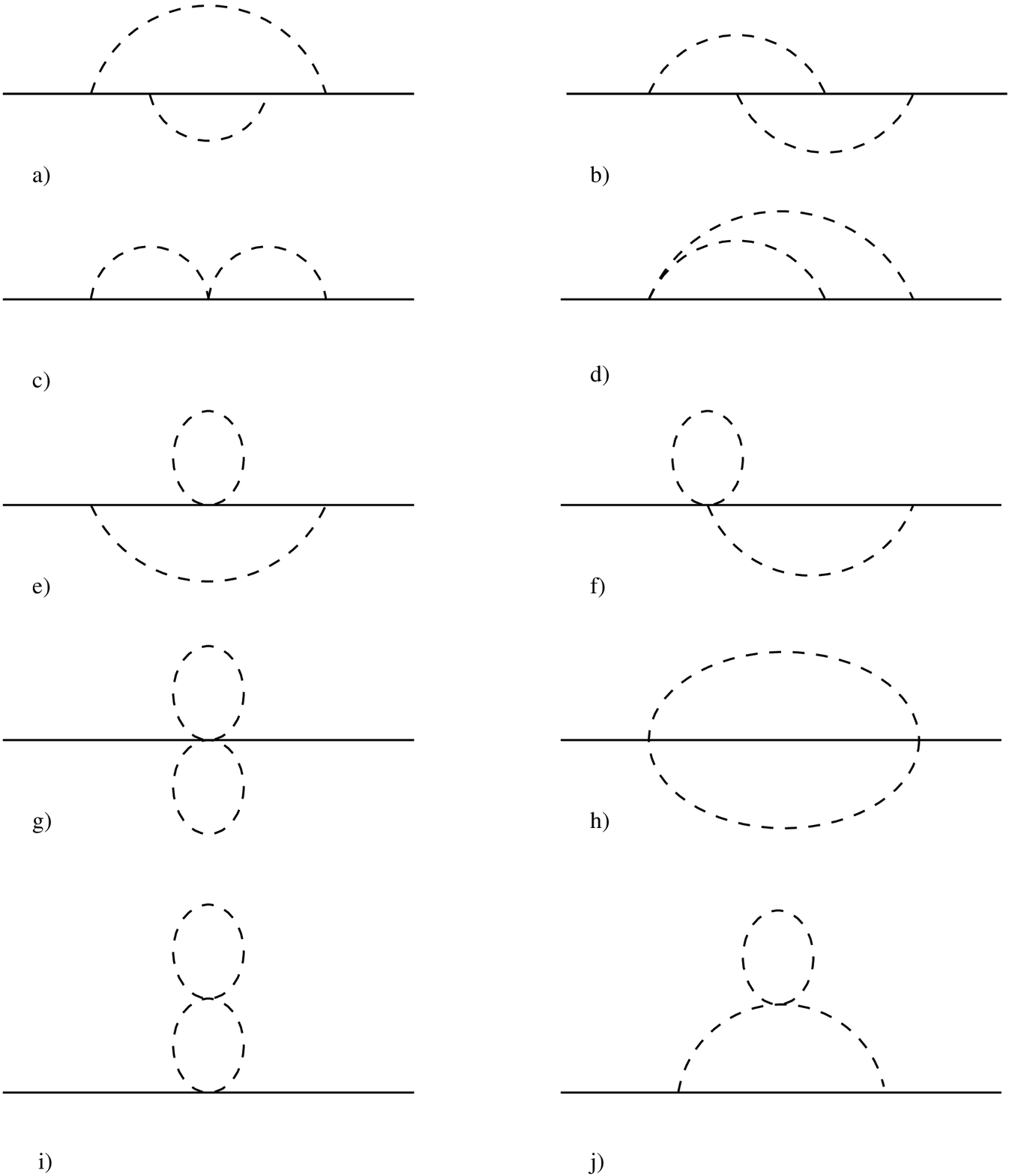}\\[2ex]
\parbox{14cm}{Figure 2: Two-loop topologies of the nucleon self-energy. Crossed
diagrams are not shown.}
\end{center}
where the coefficients $k_i$ are given by
\begin{eqnarray} 
\label{parki} 
&&k_1=-4 c_1,\quad 
k_2=-\frac{3 {\stackrel{\circ}{g_A}}^2}{32\pi F^2},\quad 
k_3=\frac{3}{32\pi^2 F^2}\left(8c_1-c_2-4 c_3
-\frac{{\stackrel{\circ}{g_A}}^2}{m}\right), 
\nonumber\\ 
&&k_4=\frac{3 {\stackrel{\circ}{g_A}}^2}{32 \pi^2 F^2 m}(1+4 c_1 m) 
+\frac{3}{128\pi^2F^2}c_2+\frac{1}{2}\alpha.
\end{eqnarray}
Here, $\alpha=-4(8 e_{38}+e_{115}+e_{116})$ is a linear combination
of ${\cal O}(q^4)$ coefficients \cite{HBLag}.
  In order to obtain an estimate for the various contributions 
of Eq.\ (\ref{mnoq4}) to the nucleon mass, we make use of 
the set of parameters $c_i$ of Ref.\ \cite{Becher:2001hv}, 
\begin{equation}
\label{parametersci} 
c_1=-0.9\,m_N^{-1},\quad c_2=2.5\, m_N^{-1},\quad
c_3=-4.2\, m_N^{-1},\quad 
c_4=2.3\, m_N^{-1} 
\end{equation} 
which were obtained from a (tree-level) fit to $\pi N$ 
scattering threshold parameters.
   Using the numerical values
\begin{displaymath}
\label{numericalvalues} 
g_A=1.267,\quad F_\pi=92.4\,\mbox{MeV},\quad 
m_N=938.3\,\mbox{MeV},\quad
M_\pi=139.6\,\mbox{MeV}, 
\end{displaymath}
\begin{center}
\includegraphics[width=9cm]{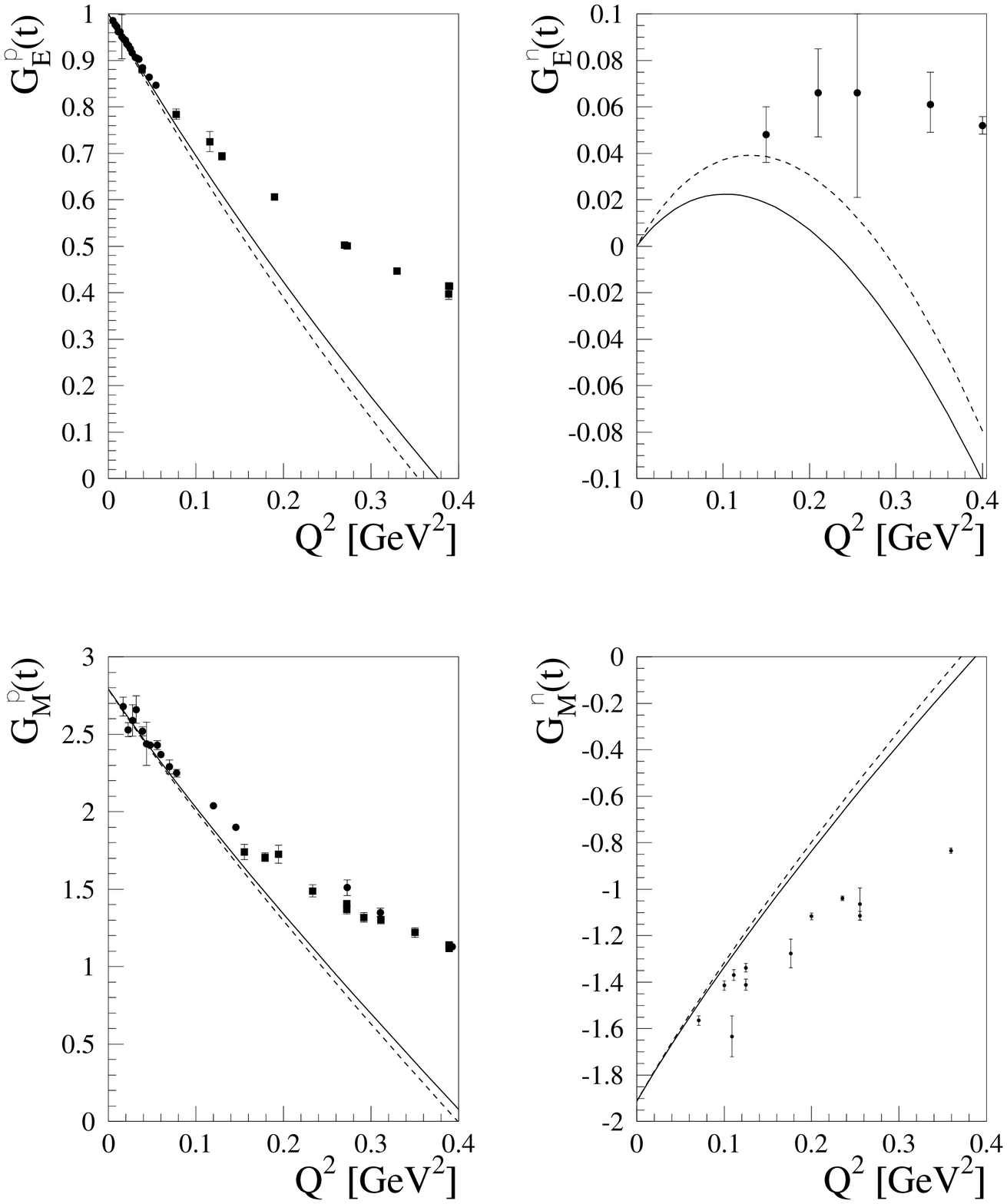}\\[1ex]
   \parbox{14cm}{Figure 3: 
The Sachs form factors of the nucleon at ${\cal O}(q^4)$. 
The solid and dashed lines refer to the results in the EOMS scheme 
\cite{Fuchs:2003ir} and
the infrared regularization \cite{Kubis:2000zd}, respectively.}
\end{center}
we obtain for the mass of nucleon in the chiral limit (at
fixed $m_s\neq 0$):
\begin{displaymath}
m=m_N-\Delta m=[938.3-74.8+15.3+4.7+1.6-2.3]\,\mbox{MeV}
=882.8\, \mbox{MeV}
\end{displaymath}
with $\Delta m=55.5\,\mbox{MeV}$. 
   Here, we have made use of an estimate for $\alpha$ obtained from
the $\sigma$ term (see Ref.\ \cite{Fuchs:2003sh} for details).
   The chiral expansion reveals a good convergence and it will be interesting 
to further study the convergence at the two-loop level \cite{Gegelia:2004}
(see Fig.~2). 

 As another example, let us consider the electromagnetic form factors 
of the nucleon which are defined via the matrix element of the electromagnetic
current operator as 
\begin{displaymath}
\langle N(p_f)\left| J^\mu(0) \right| N(p_i) \rangle=
   \bar{u}(p_f)\left[\gamma^\mu F_1^N(Q^2)+ 
\frac{i\sigma^{\mu\nu}q_\nu}{2m_N}F_2^N(Q^2) \right] u(p_i), \,\, N=p,n, 
\end{displaymath} 
where $q=p_f-p_i$ is the momentum transfer and 
$Q^2\equiv-q^2=-t \ge 0$.
   Figure 3  shows the results for the electric and
magnetic Sachs form 
factors $G_E=F_1 - Q^2/(4m_N^2) F_2$ and
$G_M= F_1 + F_2$ 
at ${\cal O}(q^4)$ in the EOMS scheme (solid lines) \cite{Fuchs:2003ir}
and the infrared renormalization (dashed lines) \cite{Kubis:2000zd}.     
    The ${\cal O}(q^4)$ results only provide a decent description up to 
$Q^2=0.1\,\mbox{GeV}^2$ and do not generate sufficient curvature
for larger values of $Q^2$.
   We conclude that the perturbation series converges, at best, slowly and
that higher-order contributions must play an important role. 
   It remains to be seen to what extent a {\em consistent} inclusion of vector
mesons \cite{Fuchs:2003sh} improves the quality of the description.

\section{Summary and Outlook}
   The EOMS scheme allows for a simple and consistent power counting in 
manifestly Lorentz-invariant baryon chiral perturbation theory 
\cite{Fuchs:2003qc}. 
   Since it can also be applied at the multi-loop level 
\cite{Gegelia:1999qt,Schindler:2003je} it will be interesting 
to consider selected two-loop examples and address the question
of convergence.   
   Moreover, the infrared renormalization of Becher and Leutwyler
has been reformulated in a form analogous to the EOMS renormalization
scheme \cite{Schindler:2003xv} and can thus also be applied to multi-loop 
diagrams with an arbitrary number of particles with arbitrary masses 
(see also Ref.\ \cite{Goity:2001ny}).
   Clearly, the method has a large potential and it will be interesting to
apply it to electromagnetic processes such as Compton scattering and
pion production.

\end{document}